\documentclass[useAMS,usenatbib]{mn2e}

\usepackage{graphicx,amssymb,color}
\usepackage[normalem]{ulem}

\title[Unstable fate of HD~131399]
{The unstable fate of the planet orbiting the A-star in the HD~131399 triple stellar system}

\author[Veras, Mustill \& G\"{a}nsicke]{
Dimitri Veras$^{1}$\thanks{E-mail: d.veras@warwick.ac.uk},
Alexander J. Mustill$^{2}$,
Boris T. G\"{a}nsicke$^{1}$
\\
$^{1}$Department of Physics, University of Warwick, Coventry CV4 7AL, UK
\\
$^{2}$Lund Observatory, Department of Astronomy and Theoretical Physics, Lund University, Box 43, SE-221 00 Lund, Sweden
}

\pubyear{2016}

\begin{document}
\label{firstpage}
\pagerange{\pageref{firstpage}--\pageref{lastpage}}
\maketitle

\begin{abstract}
Validated planet candidates need not lie on long-term stable orbits,
and instability triggered by post-main-sequence stellar evolution can generate
architectures which transport rocky material to white dwarfs, polluting them.
The giant planet HD~131399Ab orbits its parent A star at a projected separation
of about 50-100 au. The host star, HD131399A, is part of a hierarchical
triple with HD131399BC being a close binary separated by a few hundred au from the
A star. Here, we determine the fate of this system,
and find that (i) stability along the main sequence is achieved only 
for a favourable choice of parameters within the errors, and (ii) even for this
choice, in almost every instance the planet is ejected during the transition 
between the giant branch and white dwarf phases of HD~131399A. This
result provides an example of both how the free-floating planet population may be 
enhanced by similar systems, and how instability can manifest in
the polluted white dwarf progenitor population.
\end{abstract}


\section{Introduction}

Direct imaging provides an invaluable window into the outer reaches of planetary
systems \citep{bryetal2016,clagau2016a,duretal2016,regetal2016}.  Beyond about 5 au, the indirect
planet-detection techniques of Doppler radial velocity and transit photometry are 
effectively blind. Nevertheless, our Solar system and the HR 8799 system \citep{maretal2008,maretal2010} 
demonstrate that many giant planets beyond 5 au can co-exist, as well
as rocky minor planets, such as Pluto, moons like Triton, and vast belts such as the Kuiper Belt
and scattered disc.

This material in the outer reaches of planetary systems largely survives the
giant branch evolution of their parent stars \citep{veras2016a}, even amidst
dynamical excitation from the presence of a binary stellar companion 
\citep{bonver2015,hampor2016a,petmun2016} or a distant Planet Nine analogue \citep{veras2016b}.
The directly imaged $7M_{\rm Jup}$ planet orbiting the white dwarf WD 0806-661 
at a distance of approximately 2500 au \citep{luhetal2011} provides a perhaps
extreme example of the survivability of some planets orbiting evolved stars.
White dwarfs in wide binaries have in fact been used to help constrain
planet formation in the presence of a stellar companion \citep{zuckerman2014}.

The direct imaging discovery of a $4M_{\rm Jup}$ planet in the 
HD~131399 triple star system \citep{wagetal2016} -- where the planet
orbits its single parent star at a distance of about 50-100 au -- provides a 
helpful opportunity to
study long-term stability across multiple phases of stellar evolution
in a dynamically complex environment.
The planet-host, HD~131399A, is an A-star (with a mass of about $1.8M_{\odot}$),
which represents the progenitor stellar type of the white dwarfs
most commonly observed today \citep{treetal2016}. The two companion
stars (HD~131399B and HD~131399C) form a tight binary whose barycentre
orbits HD~131399A at a distance of just a few hundred au 
(see Tables \ref{SpectralTypeMass}-\ref{Orbit}).  The wide separation 
of the planet allows for ensembles of full-lifetime simulations to be carried out,
and the tightness of the HD~131399B and HD~131399C mutual orbit allows
them to be treated as a single object (see e.g. right panel of Fig. S3 of
\citealt*{wagetal2016}).

This paper explores the fate of HD~131399Ab. 
We first describe in Section 2 why studying
the long-term evolution of planetary systems is so important,
before setting up the simulations in section 3, presenting 
the results in section 4, and concluding in section 5.

\section{Importance of determining fate}

The fates of planetary systems provide unique chemical and dynamical 
constraints that directly link to their formation. 

Chemically, within
the atmospheres of white dwarfs up to 20 different metals from planetary
debris have now been measured 
\citep{gaeetal2012,juryou2014,xuetal2014,melduf2016,wiletal2016}. This debris
predominately arises from tidal break-up 
\citep{debetal2012,veretal2014a,veretal2015a,veretal2016a} 
of progenitor asteroids which have compositions that could be mapped to 
particular Solar system asteroid families (e.g. Fig. 7 of \citealt*{gaeetal2012} 
and Fig. 10 of \citealt*{wiletal2015}). In about 40 cases,
discs of this debris have been detected \citep{zucbec1987,farihi2016}.
All of the discs are dusty, and gaseous components have been detected in 
some \citep{gaeetal2006,manetal2016a}. The discs themselves
are protean, demonstrating a remarkable variability \citep{wiletal2014,xujur2014,manetal2016b},
and potential eccentricity \citep{denetal2016},
and all orbit white dwarfs which are chemically polluted. The likely disruption of an asteroid
orbiting WD 1145+017 \citep{vanetal2015,gaeetal2016,garetal2016,guretal2016,rapetal2016,veretal2016a}
is accompanied by both chemical signatures in circumstellar gas \citep{xuetal2016}
and chemical pollution within the white dwarf itself.

Dynamically, snapshots of planetary systems at different ages help piece
together their life cycles. Old main sequence systems \citep[e.g.][]{cametal2015},
dust and planets around giant branch stars \citep{bonetal2014,trietal2015,liletal2016,witetal2016},
and polluted white dwarfs at a variety of cooling ages \citep{koeetal2014,holetal2016}
all provide necessary constraints. Main-sequence planetary studies have now begun
to utilise white dwarf atmosphere chemical signatures in their planetary formation
models \citep{caretal2012,moretal2014,beretal2015,rametal2015,moretal2016,spietal2016}, 
and future missions such as {\it PLATO}
will further enable comparisons with well-constrained stellar ages \citep{veretal2015b}.   
Full-lifetime simulations incorporating all of 
the necessary physics (Fig. 2 of \citealt*{veras2016a}) have yet to be achieved, although more modest 
attempts have succeeded in modelling for multiple Gyr the mutual interactions amongst 
planets \citep{vergae2015,veretal2016b,veretal2016c}, and planets and asteroids 
\citep{musetal2016a}.  These types of simulations also help determine the viability of 
some post-main-sequence observations by testing for past stability 
\citep{musetal2013,portegieszwart2013}.

One of the most common outcomes of instability in these simulations is ejection, a 
process that contributes to the free-floating planet population.  The striking result 
that there might exist up to two giant planet free floaters for each Milky Way main sequence 
star \citep{sumetal2011} has yet to be explained by theory 
\citep{verray2012,foretal2015,wanetal2015,smuetal2016,sutfab2016} and is mitigated 
by the possibility that a fraction of the purported free-floaters are in fact
wide-orbit planets 
\citep{clagau2016b}\footnote{Nevertheless the true fraction of giant planet free floaters
-- perhaps closer to one per main sequence star -- remains high relative to predicted
occurrence rates from other exoplanet detection techniques.}. In
any case, mapping ejection prospects with architecture and time over 
different phases of stellar evolution 
\citep{veretal2011,vertou2012,musetal2014,veretal2014b,kosetal2016}  
can also be explored with the HD~131399 system.

\section{System setup}

   \subsection{Observational constraints}

HD~131399 is a triple star system with one planet (HD~131399Ab) orbiting one of the stars (HD~131399A), 
with the other two stars (HD~131399B and HD~131399C) forming a close binary \citep{wagetal2016}.
The masses and spectral types of all four objects are given in Table \ref{SpectralTypeMass}.

Because the system components have been directly imaged, the observables are projected separations,
rather than semimajor axes. The large separations imply that years or decades of observations would
be necessary to better constrain orbital properties.  Fortunately, \cite{wagetal2016} partly accomplished
this task by also utilising astrometric data dating back to 1897 \citep{gill1897}.
The result is the orbital parameters given in Table \ref{Orbit}. No orbital 
parameters were estimated for the mutual orbit of HD~131399B and HD~131399C, although \cite{wagetal2016} 
reports that the difference in their projected separations from HD~131399A is only about
7 au.

\begin{table}
\caption{Spectral types and masses of
the components of the HD~131399 system. 
All data is reproduced from Wagner et al. (2016).}
\begin{tabular}{| c | c  c  c  c}
Component  & Ab & A & B & C \\
\hline
Spectral Type   & -- & A & G & K \\
Mass   & $4\pm 1 M_{\rm Jup}$ & $1.82M_{\odot}$ & $0.96M_{\odot}$  & $0.60M_{\odot}$ \\
\end{tabular}
\label{SpectralTypeMass}
\end{table}

\begin{table}
\caption{Orbital parameters from the orbit
of HD~131399A and  HD~131399Ab ({\it top row})
and the orbit
of the barycentre of HD~131399A and HD~131399Ab
and the barycentre of HD~131399B and HD~131399C
({\it bottom row}). $a$, $e$ and $i$
refer to the semimajor axis, eccentricity and
inclination with respect to the plane of the sky. Recall that mutual inclination is defined with respect to both inclination and longitude of ascending node, meaning mutual inclinations generally differ from those assumed by simply looking at the values of $i$ here. 
All data is reproduced from Wagner et al. (2016).}
\begin{tabular}{c  c  c}
$a_{\rm pl}$  & $e_{\rm pl}$ & $i_{\rm pl}$  \\
$82_{-27}^{+23}$ au  &  $0.35 \pm 0.25$  & $40_{-20}^{+80}$ degrees \\
\hline
$a_{\rm bin}$  & $e_{\rm bin}$ & $i_{\rm bin}$  \\
$270-390$ au  &  $0.1-0.3$  & $30-70$ degrees
\end{tabular}
\label{Orbit}
\end{table}

%%%%%%%%%%%%%%%% Figure
\begin{figure*}
\centerline{
\includegraphics[width=8cm]{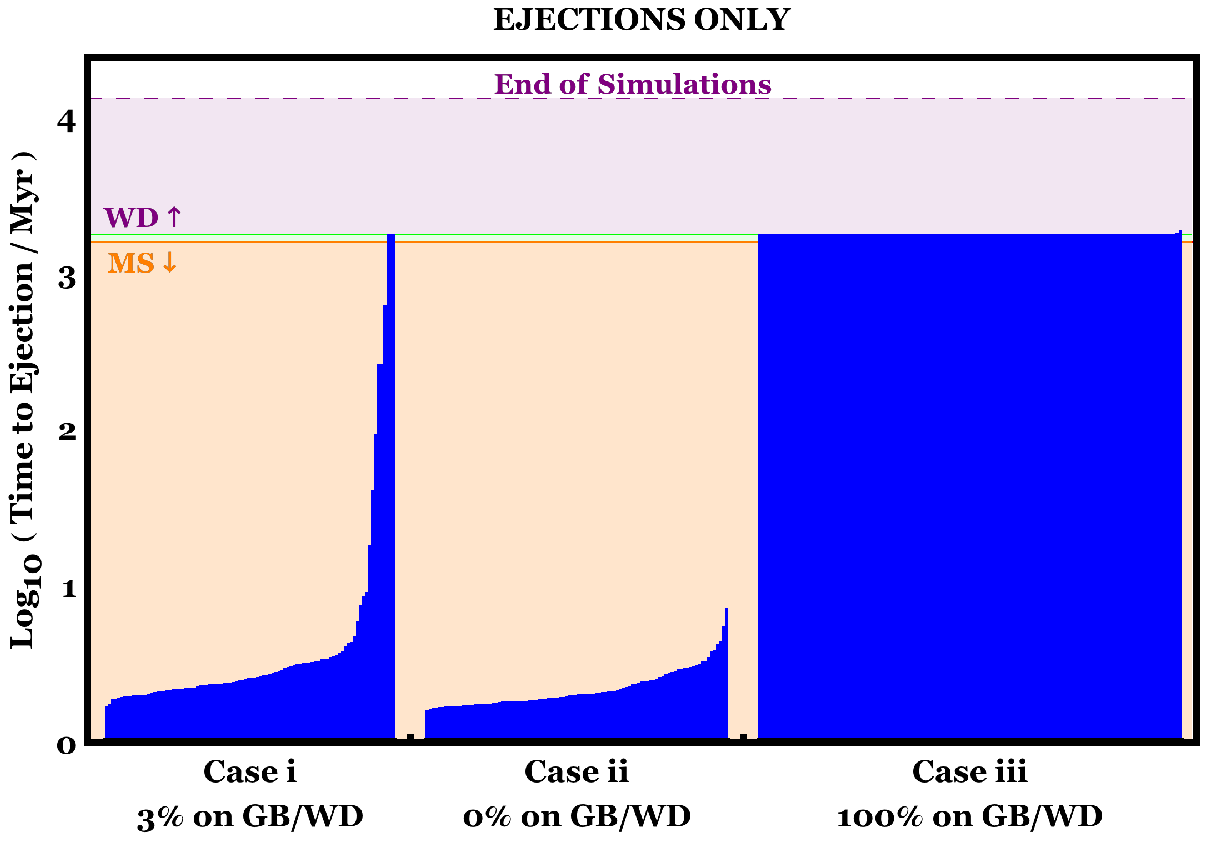}
\includegraphics[width=8cm]{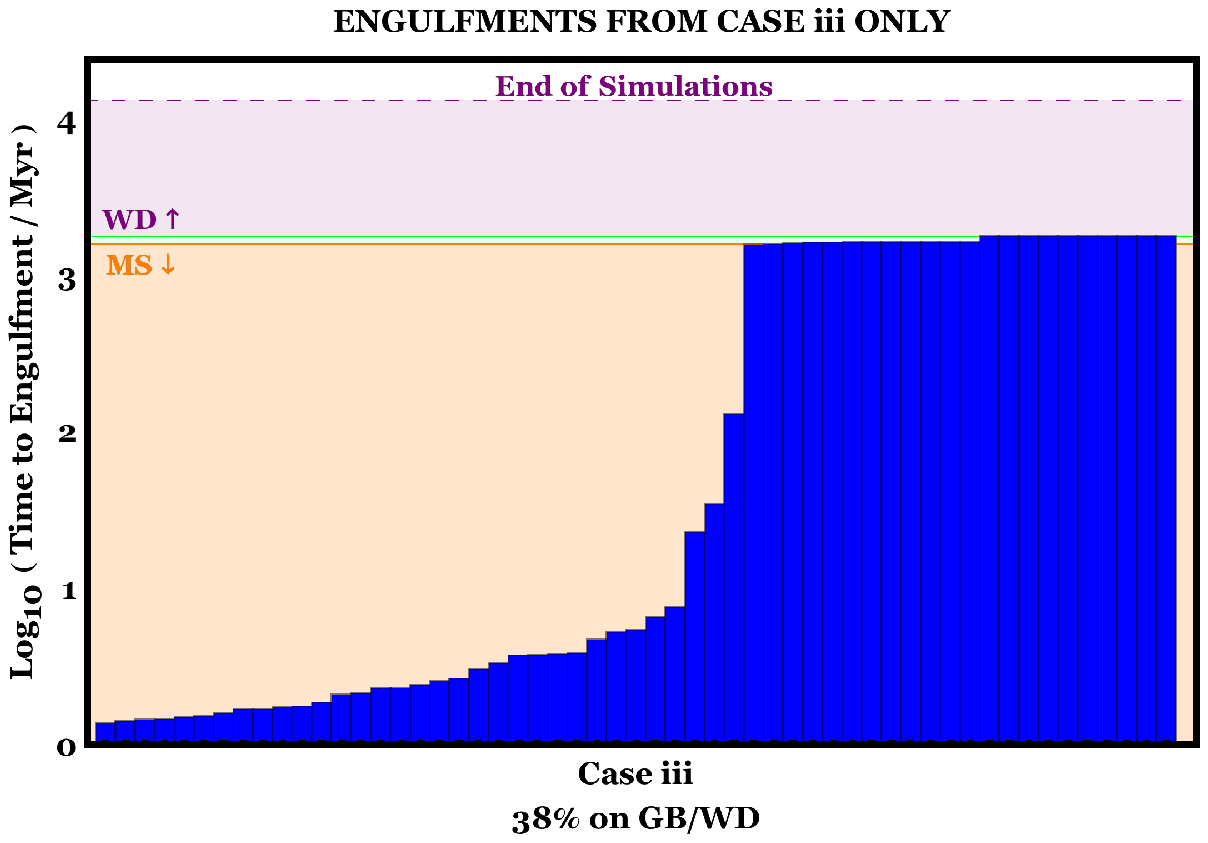}
}
\centerline{
\includegraphics[width=8cm]{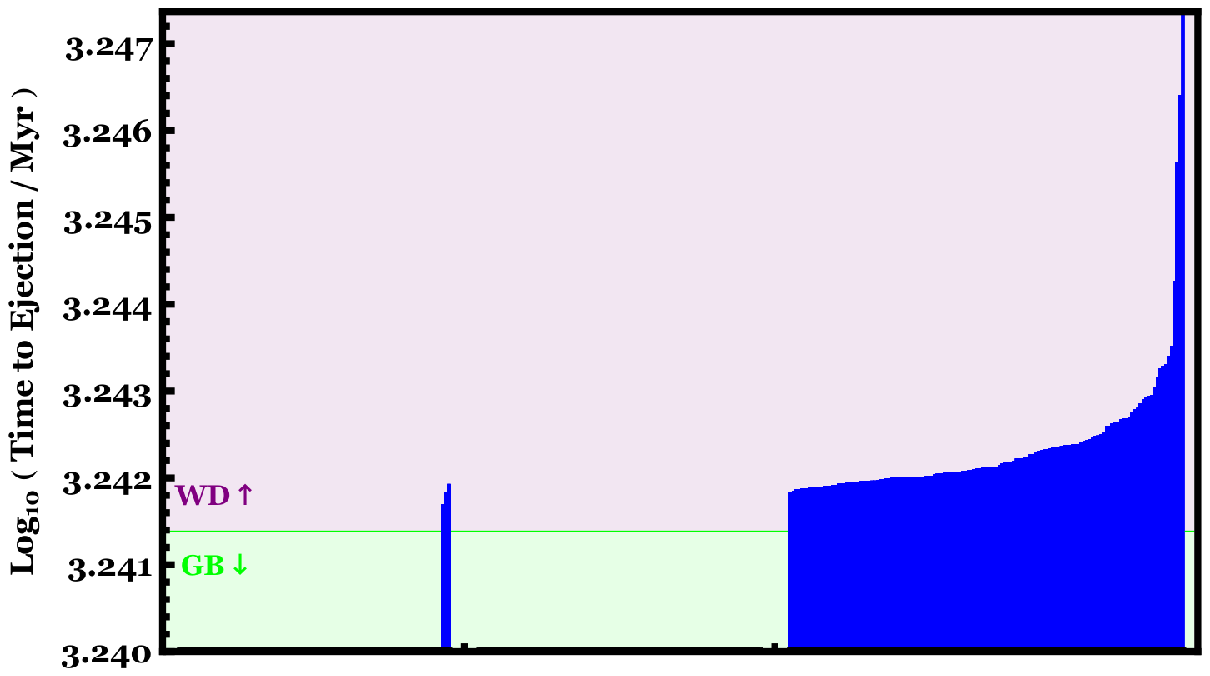}
\includegraphics[width=8cm]{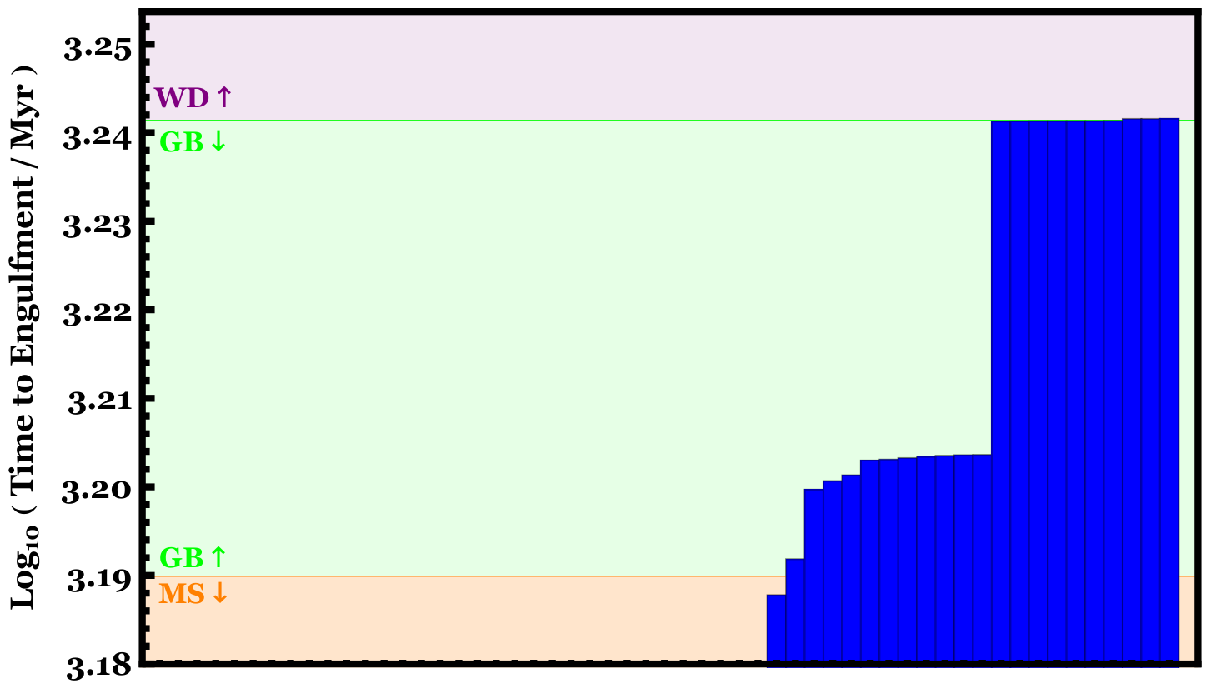}
}
\caption{
Sorted ejection times ({\it left-hand panels}) and engulfment times ({\it right-hand panels})
by case ({\it case~i: nominal parameter choice; case~ii: precarious parameter choice; 
  case~iii: guarded parameter choice} - see Sec. 3.3.1.) for the planet HD~131399Ab.
The bottom panels are zoomed-in
versions of the top panels. Each 
individual bar represents a single simulation, 
and time zero corresponds to the currently measured configuration, which is thought to be about
16 Myr old (Wagner et al. 2016). The $x$-axes give the case numbers as well as the fraction of
ejections and engulfments that occurred on either the giant branch or white dwarf phase
of HD~131399A. Different stellar phases are indicated by shaded regions: orange 
for main sequence, green for giant branch, and purple for white dwarf. The
purple dashed line indicates the ending time of the simulations. Instability occurs
in almost every simulation, and nearly-exclusively on the main sequence or during the transition
between the giant branch and white dwarf phases, depending on the parameters adopted. In the
bottom-right panel, engulfments occur predominantely at the tips of the red giant and
asymptotic giant branch phases of HD~131399A.
}
\label{unstable}
\end{figure*}
%%%%%%%%%%%%%%%% Figure

   \subsection{Our model}

This close proximity of the two non-planet host stars allow us to treat
them as a single body with a combined mass of $1.56M_{\odot}$ but with
a main sequence lifetime befitting the more massive star (HD~131399B; $0.96M_{\odot}$).
We can quantify the goodness of this approximation by considering equation 10 of \cite{hampor2016b}, which represents the ratio of Lidov-Kozai timescales of different orbits in this system. Assume orbit \#1 denotes the orbit of HD~131399Ab around HD~131399A, orbit \#2 denotes the mutual orbit of HD~131399B and HD~131399C, and orbit \#3 denotes the widest orbit, containing both orbits \#1 and \#2. Then when the equation is applied to the orbit pairs (1,3) and (2,3), for nominal parameters (Table \ref{Orbit}) and assuming a semimajor axis of 7.5~au for orbit \#2, then the ratio is about $0.03 \ll 1$. Hence, at least from a secular dynamics perspective, the binarity of stars HD~131399B and HD~131399C can be neglected.

If one assumes that HD~131399B has Solar metalicity, then the {\tt SSE} stellar evolution
code \citep{huretal2000} predicts a main sequence lifetime of about 12.8 Gyr.
In contrast, the main sequence lifetime of the $1.82M_{\odot}$ star HD 131399A
is about just 1.5 Gyr, with a shorter giant branch lifetime of 0.2 Gyr.

Therefore, we executed integrations with three bodies (HD~131399A, HD~131399Ab
and the approximated outer stellar companion) for 12.8 Gyr, treating the outer
companion as a main sequence star throughout while evolving HD~131399A through the 
main sequence, giant branch and white dwarf phases of evolution.

   \subsection{Numerical code}
We performed the simulations by using a RADAU-based integrator
within the {\it Mercury} suite \citep{chambers1999} that interpolates
stellar mass and radius changes from {\it SSE} \citep{huretal2000}.
Full details of this combined, open-source code are provided in \cite{musetal2016b}.
The code improves upon the previous incarnation, a Bulirsch-Stoer
based integrator, which was used in several previous studies and introduced
in \cite{veretal2013}. We adopted an accuracy parameter of $10^{-12}$.

       \subsubsection{Initial conditions}
We performed a total of 400 simulations, split into three cases (below) based on the
error ranges in \cite{wagetal2016}. Within each case, we fixed the mass of
HD~131399Ab to be $4M_{\rm Jup}$, and randomly sampled from a uniform distribution
(1) the arguments of pericentre, longitudes of ascending node, and mean anomalies 
of both orbits in Table \ref{Orbit}, and (2) the inclinations
across the ranges given in those tables. These values are somewhat constrained by
the observations for the binary, but not for the planet. The constraints on the
orbit of the planet are too weak to be useful, justifying this naive approach.

\begin{enumerate}

\item {\bf The nominal case; 100 simulations}: Here we adopted the nominal values of
semimajor axes and eccentricities that were given in \cite{wagetal2016}. These corresponded
to $a_{\rm pl} = 82$ au, $e_{\rm pl} = 0.35$, $a_{\rm bin} = 330$ au and $e_{\rm bin} = 0.2$.

\item {\bf The precarious case; 100 simulations}: Here we minimised the difference
between the apocentre of the planet orbit and the pericentre of the mutual stellar orbit.
These choices corresponded to $a_{\rm pl} = 105$ au, $e_{\rm pl} = 0.70$, $a_{\rm bin} = 270$ au 
and $e_{\rm bin} = 0.3$.

\item {\bf The guarded case; 200 simulations}:  Here we maximised the difference
between the apocentre of the planet orbit and the pericentre of the mutual stellar orbit.
These choices corresponded to 
$a_{\rm pl} = 55$ au, $e_{\rm pl} = 0.10$, $a_{\rm bin} = 390$ au and $e_{\rm bin} = 0.1$.

\end{enumerate}

Because the maximum apocentre of the mutual stellar orbit that we considered was $429$ au,
we have neglected effects from Galactic tides and stellar flybys across all stellar
phases \citep{veretal2014b} but adopted realistic Hill ellipsoids to model 
escape \citep{vereva2013,veretal2014c} assuming a circular Galactic orbit at 8 kpc.

Even before running these simulations, one may obtain a rough sense of the expected outcomes on the main sequence only by appealing to existing stability criteria. One well-used criterion is that from equation 1 of \cite{holwie1999}, which estimates the critical semimajor axis within which a circumstellar test particle would be come unstable in the presence of a binary stellar companion. This criterion also assumes full coplanarity, and has been shown to not be fully accurate due to limiting sampling resolution \citep[see][]{margal2016}. Nevertheless, the \cite{holwie1999} criterion yields critical semimajor axes of about 72, 50 and 99 au, for cases (i), (ii) and (iii) respectively. These values predict that the simulations in cases (i) and (ii) would largely become unstable on the main sequence, and those in case (iii) would remain stable during this phase. Such outcomes are largely borne out by the results of the simulations, which we now present.

\section{Simulation results}

The immediate and over-riding result of our simulations 
(see Fig. \ref{unstable}) is that
397 out of all 400 simulations eventually become unstable, and do
so almost exclusively on the main sequence or during the
transition between
the giant branch and white dwarf phases. 
All 200 simulations from cases (i) and (ii) become unstable.
In the `precarious case', all systems became unstable on the main
sequence, and within 10 Myr, and such that the instability is
in the form of ejection. In the `nominal case', 96\% of all systems
featured ejections, 97\% of which occurred on the main sequence.
Of the four systems which did not feature ejection, three 
showcased planet engulfment into HD~131399A and the other
engulfment into the approximated star. The engulfments into HD~131399A
all occurred after the star became a white dwarf, but
within 3 Myr of that phase change. The engulfment into the
approximated star occurred within a few Myr of the start of
the simulation, and most likely would have resulted in ejection
if the binary was resolved \citep{smuetal2016}.

The `guarded case' (case iii) has a more varied set of outcomes,
as highlighted in Fig. \ref{unstable}. Of the 197 simulations
which became unstable, 70\% were in the form of ejections,
28\% in the form of engulfment into HD~131399A, and the remainder
engulfment into the approximated star. Every ejection occurred
after the star became a white dwarf, and within 19 Myr of that moment
in all but one case (this case corresponding to
a white dwarf ``cooling age'' of 102 Myr).
Alternatively, engulfments into HD~131399A occurred at a variety
of times over all phases (right-hand panel of Fig. \ref{unstable}),
implying that the qualitative dynamics are highly sensitive to
the choice of inclinations, longitudes of ascending node, arguments
of pericentre and/or mean anomalies. However, as seen from the
bottom-right panel of Fig. \ref{unstable}, clusters of these engulfments
occur near the end of the red giant branch and asymptotic giant branch
phases of HD~131399A.  The one planet which
featured engulfment into the approximated star did so 12~Myr after
HD~131399A became a white dwarf.

The foundation of the large qualitative dynamical difference between
the guarded case and the other two cases is the initial separation
between the the planet and the approximated star.
In the nominal and precarious cases, this separation is almost always
small enough to trigger three-body gravitational scattering, and predominately ejection.
In the guarded case, the separation is larger, but in most instances
not large enough to generate a different outcome. In the other instances,
however, the planet experiences secular oscillations in eccentricity and inclination
from the approximated star. Eventually the eccentricity becomes high enough
to create a collision, especially after HD~131399A's radius is inflated
at the tips of the red giant branch and asymptotic giant branch phases.

\begin{figure}
\centerline{
\includegraphics[width=8cm]{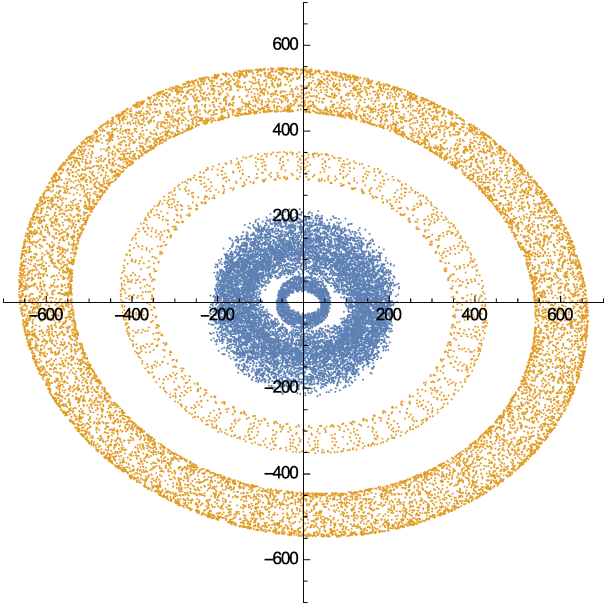}
}
\caption{
$y$-$z$ plane schematic in au of one of the only three simulations (out of 400 total)
which remained stable for 12.8 Gyr. In this instance, the initial orbital parameters were
$a_{\rm pl} = 55$ au, $e_{\rm pl} = 0.1$, $i_{\rm pl} \approx 112^{\circ}$,
$a_{\rm bin} = 390$ au, $e_{\rm bin} = 0.1$, and $i_{\rm bin} \approx 55^{\circ}$. All simulation
outputs are plotted. The blue dots represent the planet
(HD~131399Ab) and the orange dots the approximated star which emulates
the combination of HD~131399B and HD~131399C at their barycentre. The inner
and outer rings for each set of dots indicates evolution along the main sequence
and white dwarf phases, respectively. Note the extended dynamic range of
the outer ring of blue dots, which is due to a more delicate stability on the
white dwarf phase of HD~131399A.
}
\label{stable}
\end{figure}

These general statements hide the more complex dependencies on each
orbital parameter in individual systems.
This sensitivity is highlighted
by the three stable simulations, which all feature tightly
clustered initial inclinations of $i_{\rm pl} = 103^{\circ}-112^{\circ}$
and $i_{\rm bin} = 55^{\circ}-69^{\circ}$. Figure \ref{stable} presents
a schematic of the evolution of one of those simulations on the
$y-z$ plane. The outer rim of blue dots indicates that even though
the planet is stable during the white dwarf phase of 
HD~131399A, that stability is fragile.

Our results are consistent with the simulations by \cite{wagetal2016},
although the scope and particulars of each set are very different.
The most important difference is the duration of the simulations:
theirs ran for 100 Myr only on the main sequence. Their chosen orbital
parameters would fall somewhere between our nominal and guarded cases,
and they imply that all of their simulations remained stable. Stability over
100 Myr on the main sequence is an easy threshold to surpass in our guarded
case, but more difficult to achieve in the nominal case (see Fig. \ref{stable}).

%%%%%%%%%%%%%%%% Figure
%%%%%%%%%%%%%%%% Figure

\subsection{Caveats}

Our results are subject to a number of caveats. 

\begin{itemize}

\item Ejection is determined at the point where the planet leaves
the Hill ellipsoid of the system. As the axes of this ellipsoid
have a scale of the order of $10^{5}$ au, a planet may take Myrs
to technically be ejected from the system after becoming unbound
from HD~131399A. Such variations (of a few Myr) do not affect our
overall results.

\item Our code does not include tidal effects, which could 
alter the engulfment statistics. Tides have a much greater
reach than the star's physical radius (by up to a factor of a few) 
when the star expands onto the red giant branch 
\citep{villiv2009,kunetal2011,adablo2013,viletal2014} 
and asymptotic giant branch \citep{musvil2012,norspi2013,staetal2016}. 
Hence, if HD~131399Ab was perturbed into an orbit with a high-enough
eccentricity such that its pericentre was within a few au of
its parent star during one of these phases, its future evolution may
be affected. Incorporation of tides
into the code is far beyond the scope of this paper, given
their complexity, and at most they would cause a marginal change
in the instability type percentages.

\item Both HD~131399B and HD~131399C are approximated as a single star.
This approximation is good enough for our purposes, given that (i) the effects
of modelling both stars would have prevented us from simulating the system for over 1 Gyr
because of the prohibitive timestep that would be required, and (ii) their differential
effect on the planet is negligible (Fig. S3 of \citealt*{wagetal2016}).
The consequence is that engulfment into those stars is not correctly modelled,
a case we encountered only a handful of times. Further, in no instance did
we see the planet ``hop'' from HD~131399A to the approximated star \citep{kraper2012};
the dynamics of hopping instead to a tight binary might represent an 
intriguing future project.

\end{itemize}

\section{Summary}

We have determined the fate of the planet in the HD~131399 triple star system across all phases of stellar evolution
of the A-star planet host, which will become a white dwarf. The computational expense of 
our long-term (12.8 Gyr) simulations restricted our exploration to three sets of semimajor axes and eccentricities
that straddle the error bars of the observations reported in \cite{wagetal2016}. We found that the planet becomes
unstable in 397 out of 400 realisations. The instability primarily comes in the form of ejection. Whether the ejection
occurs on the planet-host's main sequence phase or during the transition between its giant branch and white dwarf phases depends
on the adopted orbital parameters. 

The strong evidence for an unstable outcome has pivotal
implications for any extant currently-undetectable smaller bodies
in the system, such as an exo-asteroid belt analogue, exo-Kuiper belt analogue,
moons or planets. Instability could
trigger excitation of belt constituents -- perturbing them into the white dwarf and polluting it,
even in the presence of a binary companion \citep{zuckerman2014} --
either from the single known planet only 
\citep{bonetal2011,debetal2012,frehan2014,antver2016} or multiple
currently unseen planets \citep{musetal2016a},
or solely due to the companion star \citep{bonver2015,hampor2016a,petmun2016}.
Liberated moons \citep{payetal2016a,payetal2016b} and multi-planet scattering within
a multiple-star system \citep{veretal2016c} can more generally contribute to active
post-main-sequence dynamical environments around systems like HD~131399.
Because the current population of metal-polluted white dwarfs largely arose from A-star 
progenitors such as HD~131399A, that star represents a notable example in the continuing effort
to appreciate the full life cycle of planetary systems.

\section*{Acknowledgements}

We thank Adrian Hamers, the referee, for his spot-on suggestions, as well as independently verifying
our claim that the binarity of HD 131399B and HD131399C does not affect our calculations. DV and BTG have received funding from the European Research Council under the European Union's Seventh Framework Programme (FP/2007-2013)/ERC Grant Agreement n. 320964 (WDTracer). AJM is supported by the Knut and Alice Wallenberg Foundation.

%\appendix


%\bsp
\label{lastpage}
\end{document}